\documentclass[11pt]{article}
\usepackage{amssymb}
\usepackage{amsfonts}
\usepackage{amsmath}

\begin{document}

\title{\textbf{General solutions of integrable cosmological models with  non-minimal coupling}}
 \author{A.Yu.~Kamenshchik$^{a,b,}$\footnote{E-mail: Alexander.Kamenshchik@bo.infn.it},  E.O.~Pozdeeva$^{c,}$\footnote{E-mail: pozdeeva@www-hep.sinp.msu.ru},  A.~Tronconi$^{a,}$\footnote{E-mail: tronconi@bo.infn.it}, G.~Venturi$^{a,}$\footnote{E-mail: giovanni.venturi@bo.infn.it}, S.Yu.~Vernov$^{c,}$\footnote{E-mail: svernov@theory.sinp.msu.ru}\\
$^a$Dipartimento di Fisica e Astronomia and INFN,\\ Via Irnerio 46, 40126 Bologna,
Italy\\
$^b$L.D. Landau Institute for Theoretical Physics of the Russian
Academy of Sciences,\\ Kosygin str.~2, 119334 Moscow, Russia\\
$^c$Skobeltsyn Institute of Nuclear Physics, Lomonosov Moscow State University,\\ Leninskie Gory~1, 119991, Moscow, Russia}
\date{ \ }
\maketitle

\begin{abstract}
We study the integrable model with minimally and non-minimally coupled scalar fields and the correspondence of their general solutions. Using the model with a minimally coupled scalar field and  a  the constant potential as an example we demonstrate the difference between the general solutions of the corresponding models in the Jordan and the Einstein frames.
\end{abstract}

\vspace*{6pt}


\section{Cosmological models with non-minimal coupling}

The observable evolution of the Universe~\cite{Planck2015}  can be described by the spatially flat
Friedmann--Lema\^{i}tre--Robertson--Walker (FLRW) background and cosmological perturbations. By this reason cosmological models with scalar fields are very
useful and play a central role in the description of the Universe. The models with the Ricci scalar multiplied by a function of the scalar field are being intensively
studied. Appearance of such a term is quite natural because quantum corrections to the effective action with minimal coupling include it~\cite{ChernikovTagirov,Callan:1970ze}. Note that the inflationary models with non-minimally coupled scalar field attract a lot of attention~\cite{nonmin-infl,HiggsInflation,EOPV2014},
because they not only do not contradict to the recent observation data~\cite{Planck2015}, but also
connect cosmology and particle physics.

We consider the model described by the following action:
\begin{equation}
\label{action} S=\int d^4 x \sqrt{-g}\left[
U(\sigma)R-\frac12g^{\mu\nu}\sigma_{,\mu}\sigma_{,\nu}-V(\sigma)\right],
\end{equation}
where $U(\phi)$ and $V(\phi)$ are differentiable functions of the scalar field $\phi$.

In a spatially flat FLRW space-time with
$ds^2 = N^2(\tau)d\tau^2 -a^2(\tau)\left(dx_1^2+dx_2^2+dx_3^2\right)$,
where $a(\tau)$ is the scale factor and $N(\tau)$ is the lapse function,
 the action (\ref{action}) leads to the following equations~\cite{KKhT,KPTVV2013}:
\begin{equation}
6Uh^2+6U'\dot{\sigma}h=\frac12\dot{\sigma}^2+N^2V\,,
\label{Fried1}
\end{equation}
\begin{equation}
 4U\dot{h}+6Uh^2+4U'\dot{\sigma}h-4Uh\frac{\dot{N}}{N}+2U''\dot{\sigma}^2
 +2U'\ddot{\sigma}-2U'\dot{\sigma}\frac{\dot{N}}{N}={} -\frac12\dot{\sigma}^2+N^2V\,,
\label{Fried2}
\end{equation}
\begin{equation}
\ddot{\sigma}+\left(3h-\frac{\dot{N}}{N}\right)\dot{\sigma} -6U'\left[\dot{h}+2h^2-h\frac{\dot{N}}{N}\right]+N^2V' = 0\,.
\label{KG}
\end{equation}
where  a ``dot'' means a derivative with respect to time and a ``prime'' means a derivative with respect to $\sigma$.
The function $h$ is the time derivative of the logarithm of the
scale factor: $h=\dot a/a$.
If we fix the lapse function $N = 1$, then $\tau$ is the cosmic time $t$ and $h$ is the Hubble parameter~$H$.

The evolutions of the Universe depends on the form of the functions $U$ and $V$. For generic functions $U$ and $V$ the system of equations
(\ref{Fried1})--(\ref{KG}) is not integrable. The number of integrable cosmological models based on scalar fields is rather limited~\cite{Fre}.
Integrable cosmological models with non-minimally coupled scalar fields correspond to integrable models with minimal coupling.
This correspondence gives a useful way to get new integrable cosmological models~\cite{KPTVV2013,KPTVV2015}. Sometimes it is easier to prove the integrability of the model with non-minimal coupling than that of the corresponding model in the Einstein frame~\cite{Boisseau:2015hqa,Pol1,Bars1}. The goal of this paper is to demonstrate that
the general solutions of the corresponding models with minimal and non-minimal coupling can be different. The reasons of this difference can be the singularities or zeroes of the functions $N(\tau)$ and $a(\tau)$, as well as zeroes of~$U(\sigma(\tau))$.

Let us perform the conformal transformation
$\tilde{g}_{\mu\nu} = 16\pi M_{\mathrm{Pl}}^{-2} U(\sigma) g_{\mu\nu}$,
where the metric in the Einstein frame is marked with a tilde, and $M_{\mathrm{Pl}}$ is the Planck mass.

After this transformation, we get a model for a minimally coupled scalar field,
described by the following action
\begin{equation}
S_E =\int d^4x\sqrt{-\tilde{g}}\left[\frac{M_{\mathrm{Pl}}^2}{16\pi}R(\tilde{g}) -
\frac{M_{\mathrm{Pl}}^2}{32\pi U}\left[1+\frac{3{U'}^2}{U}\right]\tilde{g}^{\mu\nu}\sigma_{,\mu}\sigma_{,\nu}-
\frac{
M_{\mathrm{Pl}}^4V}{256\pi^2U^2}\right]\,.
\label{actionE}
\end{equation}

Note that action (\ref{actionE}) has a singular point at $U=0$.
In this paper we consider
\begin{equation}
U_{c}(\sigma) = U_0 - \frac{\sigma^2}{12},\quad\mbox{where}\quad U_0=\frac{M_{\mathrm{Pl}}^2}{16\pi}\,.
\label{conf-coupl}
\end{equation}
i.e. we consider the case when the coupling is conformal and the standard Einstein--Hilbert term is also present. Models with the coupling function~$U_c$ are actively studied~\cite{KPTVV2013,KPTVV2015,Maciejewski:2008hj,Boisseau:2015hqa,Pol1,KPTVV2016,KTVV2013}.

In order to get the action with the standard kinetic term of the scalar field from (\ref{actionE}) we introduce a new scalar field $\phi$ such that
\begin{equation}
\frac{d\phi}{d\sigma} = \frac{\sqrt{U_0}\sqrt{U_c+3{U_c'}^2}}{U_c}=\frac{12U_0}{12U_0-\sigma^2}
\quad\Rightarrow\quad
\frac{d\sigma}{d\phi}=1-\frac{1}{12U_0}\sigma^2.
\label{scal1}
\end{equation}

Equation~(\ref{scal1}) has trivial solutions $\sigma_0=\pm\sqrt{12U_0}$ and the following nontrivial solution:
\begin{equation}
\sigma_1=\sqrt{12U_0}\tanh\left(\frac{\phi-\phi_0}{\sqrt{12U_0}}\right).
\label{connection_p}
\end{equation}
Note that functions $\sigma_1$ and $\phi$ should be real, whereas a constant $\phi_0$ can be complex. In particular, at $\phi_0=\tilde{\phi}_0+\mathrm{i}\pi/2$, we get a real solution
\begin{equation}
\sigma_2=\sqrt{12U_0}\coth\left(\frac{\phi-\tilde{\phi}_0}{\sqrt{12U_0}}\right),
\label{connection_n}
\end{equation}
if $\tilde{\phi}_0$ is a real constant. We obtain that any real solution $\phi(\tau)$ in the Einstein frame corresponds to two real solutions
$\sigma_i(\phi(\tau))$ in the Jordan frame (hereafter we consider both $\phi_0$ and $\tilde{\phi}_0$ as real numbers). Also, $U_c>0$ for any $\sigma_1$ and $U_c<0$ for any $\sigma_2$.
If  $|\phi| \rightarrow \infty$, then  $\sigma\rightarrow\sigma_0$ and, hence, $U_c\rightarrow 0$. We come to conclusion that there is no solution $\sigma(\phi(\tau))$ that crosses the value of $\sigma_0$.

On the other hand, when we consider the proper dynamics of the field $\sigma$  there is nothing that prevents it from crossing the value $\sigma = \sigma_0$, because $U_c=0$ is not a singular point of this system (\ref{Fried1})--(\ref{KG}). Indeed, we can represent this system with $U=U_c$ and $N=1$ as a dynamical system~\cite{Pozdeeva2014} and for $U=U_c$ we obtain~\cite{KPTVV2016}
\begin{equation}
\ddot\sigma={}-3H\dot\sigma -\frac{\left(12U_0-\sigma^2\right)V'+4\sigma V}{12U_0},\quad
\dot H={}-\frac{1}{12U_0}\left[2\sigma^2H^2+\left(4H\dot\sigma-V'\,\right)\sigma+2(\dot\sigma)^2\right].
\label{FOSEQU}
\end{equation}
The equations (\ref{FOSEQU}) can have a solution such that $U_c>0$ at some moment and $U_c<0$ at another moment and that can not be found on using the Jordan--Einstein frame correspondence. To clarify this statement we consider the well-known cosmological model with a minimally coupled scalar field and a positive cosmological constant. Note that models with negative or zero cosmological constant that have been considered in detail in~\cite{KPTVV2016} confirm this statement as well.

\section{The model with minimal coupling and a positive cosmological term}
Let us consider the cosmological model with a minimally coupled scalar field and a constant potential.
The corresponding equations of motion are Eqs.~(\ref{Fried1})--(\ref{KG}) with $U=U_0$ and $V=\Lambda > 0$.
At $N\equiv 1$  we obtain the following system ($\tau=\tilde{t}$ is a cosmic time):
\begin{equation}
6U_0\tilde{H}^2=\frac12\dot{\phi}^2+\Lambda,
\label{Fried-mass}
\end{equation}
\begin{equation}
\dot{\tilde{H}}+3\tilde{H}^2-\frac{\Lambda}{2U_0}=0,
\label{Fried300m}
\end{equation}
\begin{equation}
\ddot{\phi}+3\tilde{H}\dot{\phi}=0.
\label{KG-mass}
\end{equation}
To emphasise that we consider the Einstein frame we denote the scalar field as $\phi$.
A non-constant Hubble parameter $\tilde{H}$ that satisfies Eq.~(\ref{Fried300m}) is either
\begin{equation*}
\tilde{H}_{-}(\tilde{t}) =\frac{\sqrt{\Lambda}}{\sqrt{6U_0}}\tanh\left(\frac{\sqrt{6\Lambda U_0}}{2U_0}(\tilde{t}-\tilde{t}_1)\right),
\quad \mbox{or} \quad
\tilde{H}_{+}(\tilde{t}) =\frac{\sqrt{\Lambda}}{\sqrt{6U_0}}\coth\left(\frac{\sqrt{6\Lambda U_0}}{2U_0}(\tilde{t}-\tilde{t}_1)\right).
\end{equation*}
The choice of the solution is fixed by Eq.~(\ref{Fried-mass}).
If $\phi$ is real, then  only the solution $H_{+}$ is possible.
From Eq.~(\ref{KG-mass}) for $\tilde{H}=\tilde{H}_{+}(\tilde{t})$ we get
\begin{equation}
\phi=\frac{2}{3}\sqrt{3U_0}\left(\ln\left[\coth\left(\frac{\sqrt{6\Lambda}}{4\sqrt{U_0}}(\tilde{t}-\tilde{t}_1)\right)\right]+\phi_0\right),
\end{equation}
where $\tilde{t}_1$ and $\phi_0$  are integration constants. Using relations (\ref{connection_p}) and  (\ref{connection_n}), we obtain $\sigma$ as a function of the cosmic time in the Einstein frame $\tilde{t}$:
\begin{equation}
\sigma_1 = \sqrt{12U_0}\tanh\left[ \phi(\tilde{t})\right],
\qquad
\sigma_2 = \sqrt{12U_0}\coth\left[ \phi(\tilde{t})\right].
\end{equation}

At first glance we have found the general solution in the Jordan frame, because $\phi(\tilde{t})$ is the general solution in the Einstein frame. However, $|\sigma_1(\tilde{t})|<\sqrt{12U_0}$ at all $\tilde{t}$, therefore, we have  solutions that correspond to positive $U_c(\sigma)$ only. Analogously the solution $\sigma_2$ corresponds to $U_c(\sigma)<0$ for all $\sigma_2(\phi(\tilde{t}))$.
At the same time there exist exact solutions for which $U_c(\sigma)$ changes sign.

 The corresponding model with $U=U_c$ and $V=\Lambda U_c^2/U_0^2$ is described by the following system:
\begin{equation}
6U_cH^2-\sigma\dot{\sigma}H=\frac12\dot{\sigma}^2+\frac{\Lambda}{U_0^2}U_c^2,\qquad
\dot{H}+2H^2=\frac{\Lambda}{3U_0^2}U_c,\qquad\ddot{\sigma}={}-3H\dot{\sigma}\,,
\label{Fried2001}
\end{equation}
obtained from Eq.~(\ref{Fried1}) with $N\equiv 1$ and system~(\ref{FOSEQU}). 

In the case of a positive $\Lambda$ system~(\ref{Fried2001}) has the following particular solution in terms of elementary functions:
\begin{equation}
\label{sigmaHp}
\sigma(t)={}\pm\frac{6\sqrt{U_0}}{\sqrt{36U_0e^{-72U_0C_2(t-t_1)}-1}},
\qquad    H(t)={}-\frac{24C_2U_0 \left(18U_0e^{-72 U_0C_2(t-t_1)}+1\right)}{36U_0e^{-72 U_0C_2(t-t_1)}-1},
\end{equation}
where $t$ is the cosmic time in the Jordan frame, the parameter $t_1$ is arbitrary and the parameter $C_2$ is defined by the following relation $\Lambda=864C_2^2U_0^3$. It is evident that the function
$U_c$ changes its sign at $t=\ln(9U_0)/(72U_0C_2)$.

Note that the system~(\ref{Fried2001}) is integrable. The general solution for this model with an arbitrary $\Lambda$ has been found in quadratures in~\cite{KPTVV2016}, where solutions at which the function $U_c$ changes its sign have been used to describe the crossing of singularities in the Einstein frame.

\section{Conclusions}

 In this paper we concentrate on the problems of the construction of general solutions of the cosmological models. We have shown that the knowledge of the general solution in the Einstein frame does not guarantee the knowledge of the general solution for the corresponding model in the Jordan frame. At the same time this knowledge help to find the general solutions solving equations in the Jordan frame~\cite{KPTVV2013}.  The similar problem arises even if we restrict ourselves to considering models with minimal coupling only. Indeed the standard way to get the general solution includes the choice of some suitable function $N(\tau)$ that allows to integrate equations. So, one obtains the general solution in parametric time, that can be different from the general solution in the cosmic time.

The
A.K. research was partially supported by the RFBR grant 14-02-00894.
 Research of E.P. is supported in part by grant MK-7835.2016.2  of the President of Russian Federation. Research of S.V. is supported in part by the RFBR grant 14-01-00707.

\end{document}